\documentclass[preprint,12pt]{elsarticle}

\usepackage{amssymb}
\usepackage{graphicx,amsmath,mhchem}
\usepackage{multirow}
\usepackage{subcaption}
\usepackage[english]{babel}
\usepackage{tabularx,ragged2e,booktabs,caption}
\usepackage[usenames]{color}

\newcommand{\change}{\textcolor{black}}
\newcommand{\C}{$^\circ$C}
\journal{The Journal of Colloid and Interface Science}

\begin{document}

\begin{frontmatter}



\title{Inferring Pore Connectivity from Sorption Hysteresis in Multiscale Porous Media}


\author[label1]{Matthew B. Pinson}
\author[label1]{Tingtao Zhou}
\author[label2]{Hamlin M. Jennings\fnref{authorfootnote}}
\fntext[authorfootnote]{Deceased in July 2015. R.I.P.}
\author[label3]{Martin Z. Bazant}

\address[label1]{Department of Physics, Massachusetts Institute of Technology}
\address[label2]{Department of Civil and Environmental Engineering, Massachusetts Institute of Technology}
\address[label3]{Department of Chemical Engineering and Mathematics, Massachusetts Institute of Technology}

\begin{abstract}
\change{
{\it Hypothesis}\\
Vapor adsorption experiments are widely used to assess pore size distributions, but the large hysteresis sometimes observed between sorption and desorption isotherms remains difficult to interpret.  Such hysteresis is influenced pore network connectivity, which has previously been modeled by percolation on infinite lattices. Our hypothesis is that percolation occurs instead through finite networks of micropores connecting accessible macropores, always exposed to the outside environment.
\\
\\{\it Theory}\\
We derive a general formula for sorption/desorption isotherms that introduces a simple measure of hierarchical pore connectivity -- the fraction $f$ of always exposed pores.  The model thus accounts for ``small world" connections in finite-size percolation, while also incorporating other hysteresis mechanisms, in single-pore filling, liquid insertion into the solid matrix, and cavitation. \\
\\
{\it Findings}\\
Our formula is able to fit and interpret both primary and scanning sorption/desorption isotherms for a variety of adsorbates (noble gases, water, and organics)  and porous materials (cement pastes, dental enamels, porous glasses, carbon black and nanotubes), including cases with broad pore-size distributions and large hysteresis. It allows quantification of the prevalence of percolating macropores in the material, even though these pores are never filled during the sorption experiments.  
A distinct bump in sorption isotherms is explained as a lowering of the barrier to nucleation of the vapor phase with a universal temperature scaling.  
}
\end{abstract}


\end{frontmatter}



\section{Introduction}
The properties of a porous material, such as strength, permeability and hydrodynamic dispersion, depend on the structure of both the solid matrix and the pore space. \change{ Although various standard experimental methods are available~\cite{lowell2012characterization}, the geometry and topology} of the pore network can be difficult to characterize experimentally, especially due to the wide range of length scales, which can extend below one nanometer. A standard characterization approach is vapor adsorption, in which the partial pressure of an adsorbate is varied and the sorbed mass (or the intensity of a nuclear magnetic resonance signal \citep{Muller2013}) measured. In principle, such measurements enable quantification of the internal surface area, the pore size distribution (PSD), and to some extent the connectivity of the pore network.

One important aspect in the interpretation of sorption isotherms is the widely-observed hysteresis between adsorption and desorption isotherms \citep{Sing1985}, and any valid model of sorption must account for this hysteresis. Analytical and numerical models have been developed \citep{Cohan1938, Mason1982, Mason1983, Tanev1993, Liu1993, Rojas2002, Ravikovitch2002}, where several mechanisms of hysteresis on different scales have been identified: 1) chemical effects on the molecular scale 2) single pore hysteresis on the scale of individual pores and 3) pore-blocking or ink-bottle effect on the scale of pore networks. When bottleneck sizes are smaller than a certain critical value, cavitation happens before the bottlenecks become empty, and modifies the shape of the hysteresis loop~\citep{fan2011cavitation, thommes2015physisorption}. \change{In this article we are mainly concerned with the case where the solid matrix of the porous media is fully wetted by the adsorbate, thus minimizing the importance of contact angle hysteresis \citep{giesche2006mercury,caro1961physical,salmas2001mercury,
lowell2012characterization}. This mechanism is important in non-wetting solid-liquid contacts such as mercury porosimetry, though it still does not explain all the hysteresis in that case\citep{giesche2006mercury,gu2018microscopic}.}

We extend the existing infinite-lattice percolation models~\citep{Mason1982, Mason1983, Mason1988, liu1993analysis, Liu1994, neimark1991percolation} in combination with single pore hysteresis and cavitation, and propose a new parameter $f$, characterizing always-exposed large pores and representing the finite-size effect of samples. We show that this parameter varies among different types of hierarchical porous materials.
We first review different hysteresis mechanisms and identify their importance in different types of hysteresis. Of particular interest are multiscale materials, where the PSD is very wide and percolation effects are beyond the explanation of a single threshold size. We show that sorption hysteresis in such materials is heavily influenced by network effects, where $f$, in addition to the coordination number $z$, plays an important role. The partial pressure of cavitaion inferred from the sorption isotherm hysteresis shows the trend predicted by theoretical models. The inferred order parameter $f$, together with PSD can be used in further material modeling. For instance, a better understanding of the prevalence of large and small pores and how they are connected will be beneficial to the modeling of transport in porous media \citep{Dullien1992}, since transport processes are much faster in larger pores. Such transport modeling is applicable to a wide variety of problems, including creep and corrosion in concrete \citep{Bertolini2004}, capture of carbon dioxide in sorbents such as zeolites or activated carbon \citep{Choi2009}, and shale gas extraction \citep{Javadpour2007}.

\section{Single pore hysteresis}
\subsection{Surface adsorption}
One mode by which a gas can be sorbed by a mesoporous material is adsorption on pore walls. Numerous isotherms have been proposed to describe the average thickness of the adsorbed layer as a function of partial pressure, of which the Langmuir \citep{Langmuir1917} and BET \citep{Brunauer1938} equations are best known. The Langmuir equation is based on the assumption that only a monolayer can adsorb, while the BET equation uses the opposite assumption that an infinite number of layers can adsorb, with no lateral interaction between adsorbed molecules. Although adsorption is likely to fall somewhere between these extremes, we use the BET equation for simplicity. The BET equation \citep{Brunauer1938} gives the average thickness of the adsorbed layer, in units of a monolayer, as
\begin{equation}
t_m(h) = \frac{c h}{[(c-1)h+1](1-h)}. \label{BET}
\end{equation}
Here $h$ is the relative pressure of the sorbate, i.e., its partial pressure divided by its saturation pressure. The constant $c$, usually determined empirically, is related to the strength of the interaction between the sorbate and the solid surface.

The validity of the BET equation has been questioned due to its neglect of lateral forces between adsorbed molecules within each layer \citep{Sing2001,SeriLevy1993,Bazant2012a}. However, it is successfully and very widely used to explain experimental results \citep{Nelsen1958,Bhambhani1972,Branton1993}. It may be that, although lateral forces are present, they mostly act to reduce variations in layer thickness without substantially impacting the average adsorbed quantity. Also, interactions between the solid surface and layers above the first may increase the number of layers in which lateral interactions are unimportant. We model surface adsorption by the BET equation in this work and reasonable results are obtained.

\subsection{Pore filling} \label{filling}
\begin{figure*}
\centering
\includegraphics[width=\textwidth]{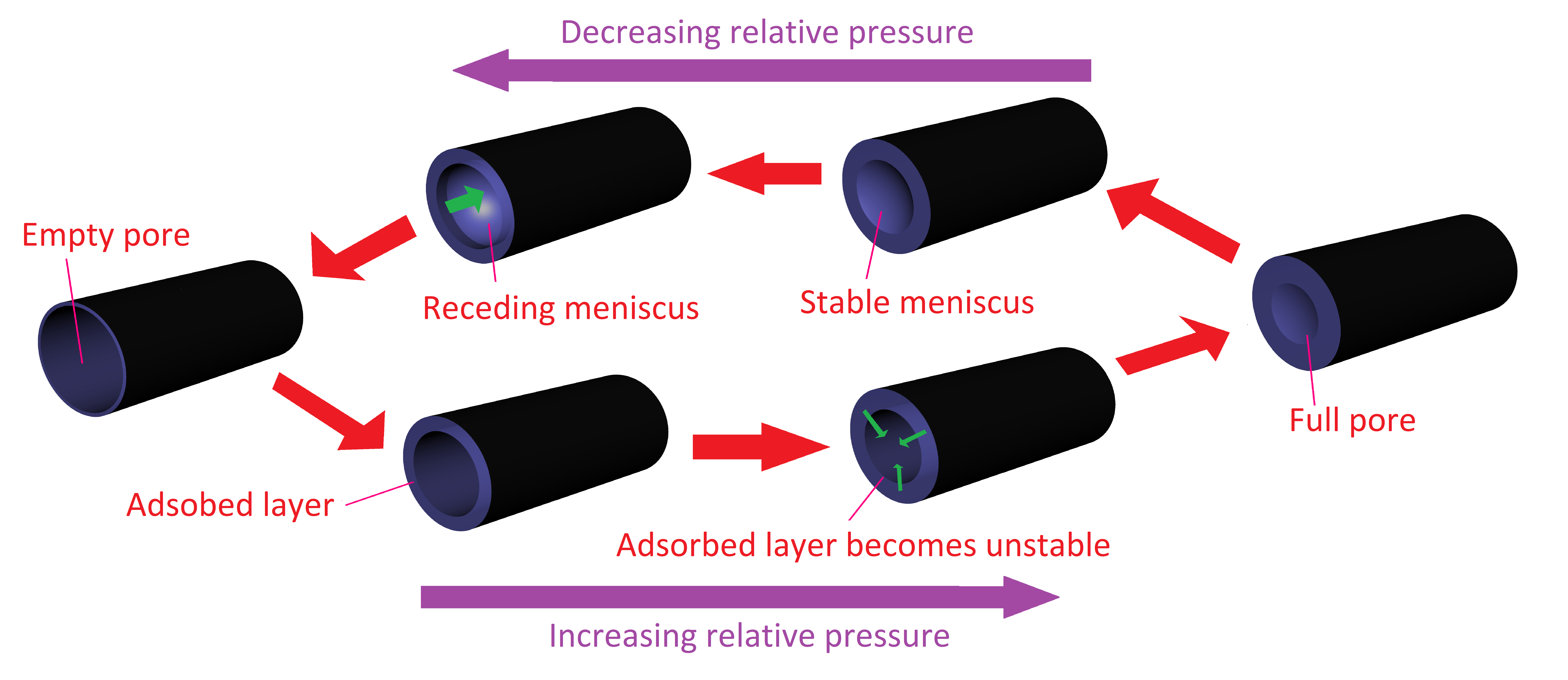}
\caption{A schematic illustration of radial filling during wetting (bottom) and axial emptying during drying (top). \label{fig:single}}
\end{figure*}

In addition to adsorption on pore walls, the sorbate can condense within the mesopores. The relationship between a radius $r$ of a liquid-vapor interface located within a pore and the relative pressure at which the full and empty (we refer to a pore containing no condensed fluid as empty, although it still has adsorbed fluid on its pore walls) states are in equilibrium is given by the Kelvin equation~\citep{barrett1951determination,thomson1872kelvinequation, stifter2000theoretical}:
\begin{equation}
\ln h = \frac{\alpha\gamma a^3}{rkT}.
\end{equation}
Here $\gamma$ is the liquid-vapor surface tension, $a^3$ is the average volume per molecule in the condensed state, $k$ is the Boltzmann constant and $T$ is temperature. The use of the Kelvin equation relies on the assumption that a continuum treatment of water is valid down to small length scales, as suggested by the explanation of freezing point suppression by an analogous continuum equation \citep{Shimizu2014}. 
The prefactor $\alpha$ accounts for the shape of the liquid-vapor interface. For example, in cylindrical pores, during adsorption the liquid film grows radially inward but during desorption the liquid-vapor meniscus recedes along the cylinder axis, causing a different prefactor ($\alpha=1$ for adsorption, $\alpha=2$ for desorption) and hence hysteresis even for a single cylindrical pore \citep{Cohan1938,Broekhoff1967,Celestini1997}. Figure \ref{fig:single} illustrates these processes. The situation of a slit pore is even more extreme: the mean radius of curvature when the pore is full is equal to the distance between the adsorbed layers on the pore walls, but since there is no pre-existing curvature the pore will not fill until the adsorbed layers grow so thick that they touch. In realistic material structures the Kelvin prefactor difference between adsorption and desorption will be further influenced by 3D geometry.

\begin{figure}
\centering
\begin{subfigure}{0.48\textwidth}
\includegraphics[width=0.98\textwidth]{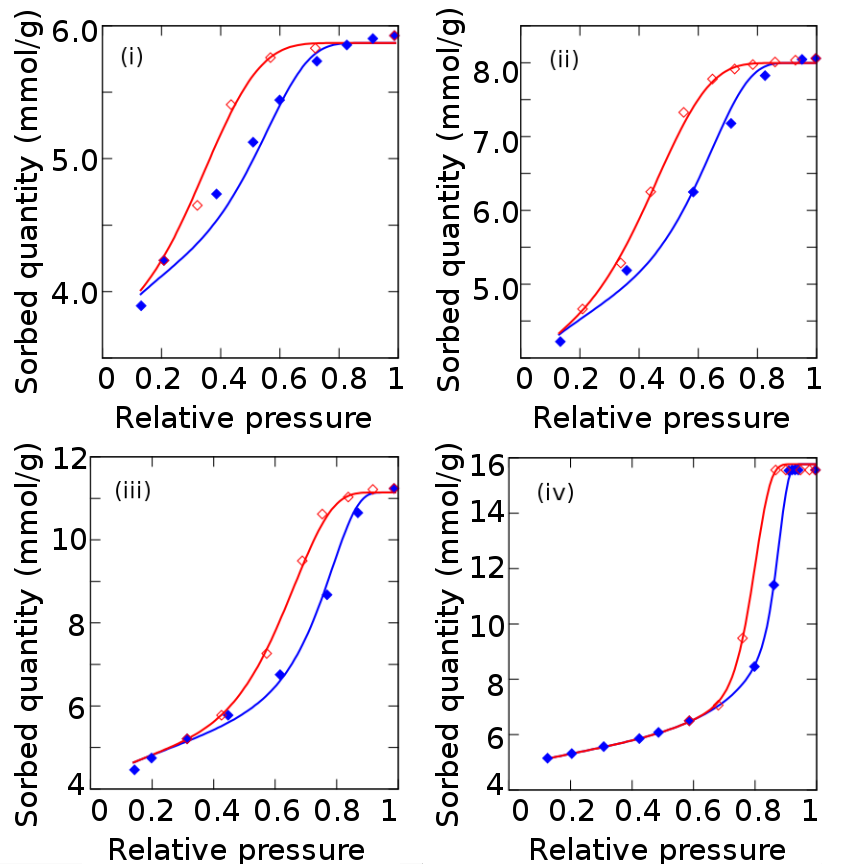}
\subcaption{ \label{machinfigure}}
\end{subfigure}
\begin{subfigure}{0.48\textwidth}
\centering
\includegraphics[width=0.98\textwidth]{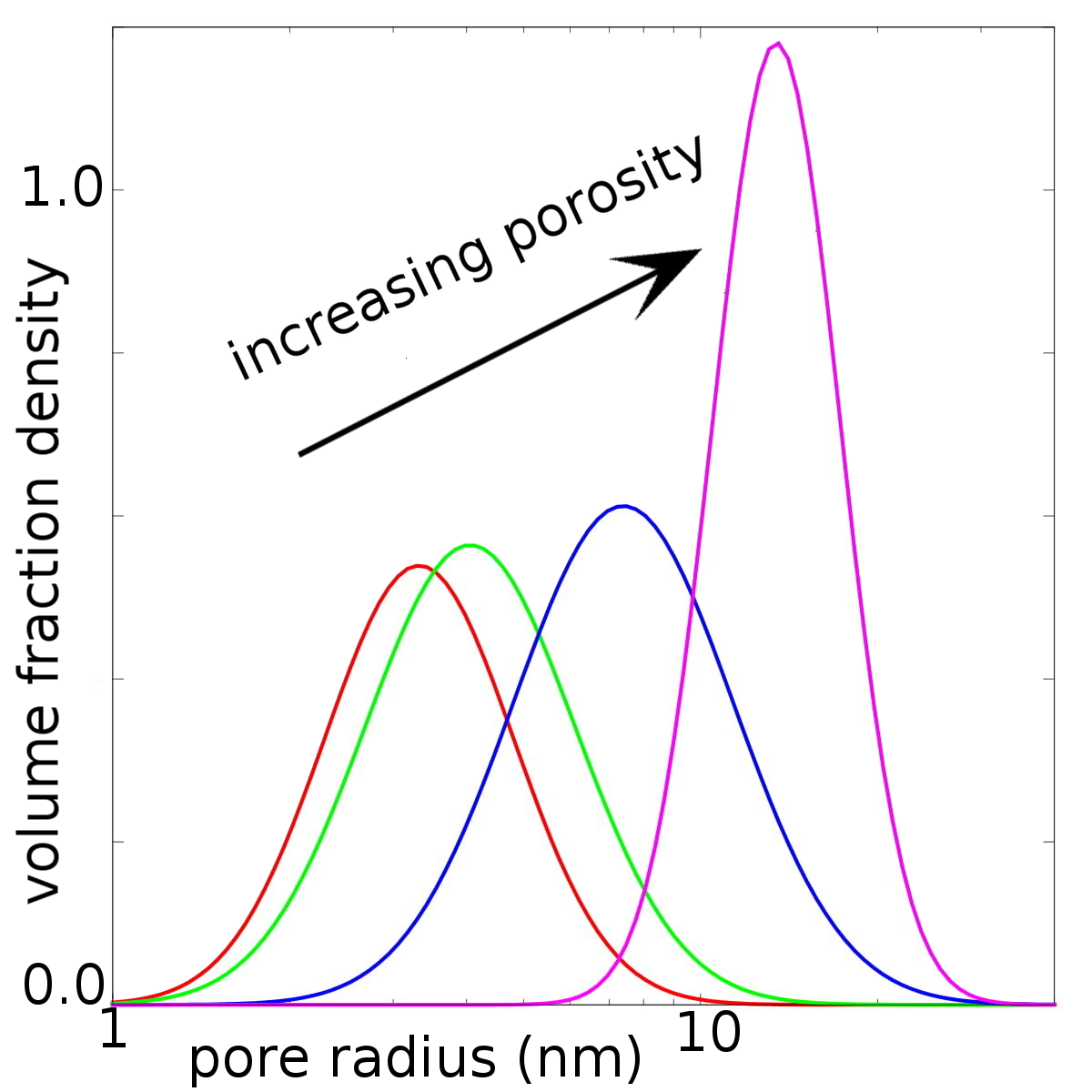}
\subcaption{ \label{machinPSD}}
\end{subfigure}
\caption{Sorption hysteresis explained by single pore filling in  of dichlorofluromethane in carbon black~\citep{Carman1951}. (a) Experimental sorption isotherms (points) for wetting (blue) and drying (red). Subfigures with porosity of (i) 0.50, (ii) 0.58, (iii) 0.66 and (iv) 0.74. Lines are the analytical results of single pore hysteresis. (b) Pore size distributions obtained by assuming cylindrical pores with only single pore filling hysteresis.
\label{fig:machin-results}}
\end{figure}

This mechanism is sufficient to describe materials with simple pore structure where network effects are negligible, such as dichlorofluromethane in plugs of carbon black \citep{Carman1951}, shown in Fig. \ref{machinfigure}. 
\change
{During drying, the cylindrical pore starts as fully saturated, and its meniscus resembles a hemisphere which has non-zero Gaussian curvature. Only when the relative pressure goes below a critical value will the meniscus start to recede and the pore empty. For an ensemble of different sizes of independent pores, given relative pressure $h$, pores larger than a critical radius will have unstable meniscus and become empty. We denote this critical radius as $r_{eq}$, defined by }
\begin{equation}
r_{eq}(h)= \frac{2\gamma a^3}{kT\ln h}+at_m(h)
\end{equation}
\change{
We denote it as the ``equilibrium desorption radius'' to emphasize that the liquid inside is in direct contact with the external vapor phase through the meniscus interface, so the meniscus interface is free to propagate or recede and eventually arrives at an ``equilibrium'' spatial position. This is in contrast to the situation where the pore is blocked by other smaller pores. In that case its meniscus interface can be held at the pore boundary instead of receding and disappearing, which would happen if it were not blocked. We will elaborate again on this difference in Section \ref{sec:network-effects}. }
If the distribution of pore radii $r$, by pore volume, is $v(r)$, the sorbed volume at relative pressure $h$ during drying is
\begin{equation}
m_d(h) = \int_0^{r_{eq}(h)}v(r)~dr+\int_{r_{eq}(h)}^\infty\frac{r^2-[at_m(h)]^2}{r^2}v(r)~dr, \label{desorbingmass}
\end{equation}
\change{
The first term accounts for all pores small enough to already be filled at $h$, the second term for the large pores that remain empty but contribute through their surface layer adsorption.
}

\change{
During wetting, a pore remains empty until the relative pressure rises above a threshold which we refer to as the filling pressure. Given $h$, all pores smaller than $r_{fill}(h)$ will be full:
\begin{equation}
r_{fill}(h) = \frac{\gamma a^3}{kT\ln h}+at_m(h)
\end{equation}.
Notice the prefactor differs by a factor of 2 between $r_{eq}$ and $r_{fill}$ for cylindrical pores, hence hysteresis is produced. This hysteresis arises from differences in the shape of the meniscus during filling and emptying.}
The total sorbed liquid volume during wetting is then
\begin{equation}
m_w(h) = \int_0^{r_{fill}(h)}v(r)~dr+\int_{r_{fill}(h)}^\infty\frac{r^2-[at_m(h)]^2}{r^2}v(r)~dr, \label{adsorbingmass}
\end{equation}
\change{
The meaning of the two terms are similar to the drying case, except that the critical radius $r_{fill}$ is different from $r_{eq}$.
}

\section{Insertion within the ``solid''}
Equations \ref{desorbingmass} and \ref{adsorbingmass} describe sorption in mesopores. Many porous materials possess a hierarchical pore structure, and sorbate can be found not only in the mesopores, but also in even smaller spaces within what would be considered the ``solid'' part of the structure \citep{Richardson2004} \citep{Richardson2008}. Simple models that treat the sorbate as a condensing fluid are not applicable to these very small (approximately one nanometer or smaller) spaces, because they neglect strong chemical interactions with the sorbent material, as well as effects arising due to the discreteness of the sorbate molecules \citep{Bazant2012}. One approach is to use various atomistic simulation techniques \citep{Bonnaud2010}. We note that in a material such as cement paste in which inserted water (known as interlayer water \citep{Kalousek1954}) is particularly important, the inserted water content as a function of relative humidity is well approximated by assuming no loss of water on drying until 15\% relative humidity is reached, linear loss of water below this, and linear reinsertion on wetting \citep{Feldman1973, Jennings2008, Muller2013a, bonnaud2012thermodynamics}. Thus we only treat the amount of inserted water at complete saturation as a unknown parameter in this work.

\section{Network effects}
\label{sec:network-effects}
\begin{figure*}
\centering
\includegraphics[width=\textwidth]{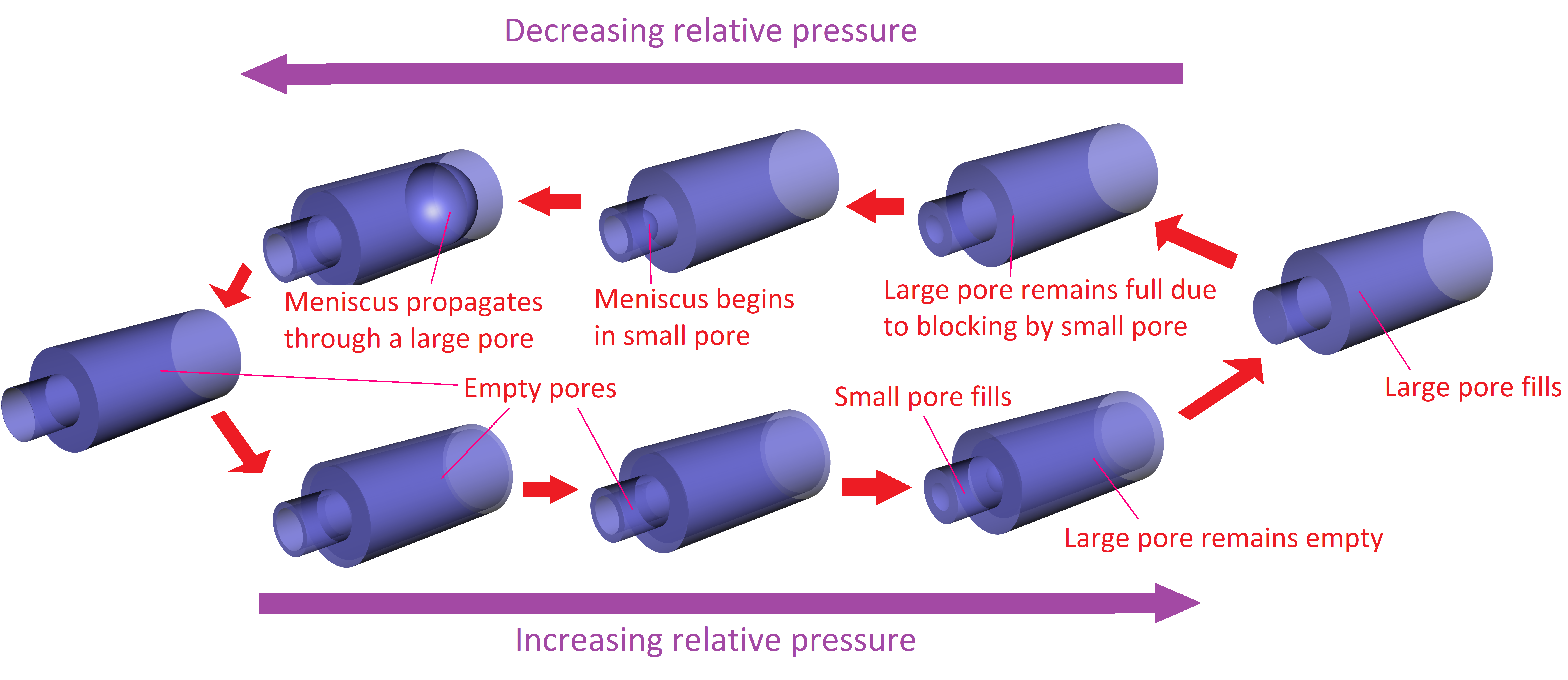}
\caption{Illustration of a system comprising a large pore whose only path to the atmosphere passes through a small pore, exposed to a cycle of relative pressure. The large pore remains full until the small pore empties and provides a meniscus where the liquid-vapor transition can be nucleated. \label{fig:2pores}}
\end{figure*}

The single pore hysteresis mechanism, accounting for hysteresis due to differences in the shape of the liquid-vapor interface in an isolated pore on wetting and drying, can accurately describe sorption in particular porous materials, with nearly independent pores. In a system with a wide range of pore sizes and good pore connectivity, network effects can act to broaden hysteresis. The additional hysteresis is due to some pores remaining full below the relative pressure at which the empty state is thermodynamically favored, because they lack the connection with the vapor phase that is necessary to nucleate the liquid to vapor transition. This is frequently known as ``pore blocking'' \citep{Tanev1993, Ravikovitch2002, thommes2015physisorption}, though it is worth noticing that the effect is not based on kinetics but purely on consideration of local equilibrium. Figure \ref{fig:2pores} illustrates this effect for a toy system comprising two cylindrical pores.

Various models of this hysteresis have been developed, applicable to certain types of pore network but not universally. Mason \citep{Mason1983} proposed a model of pores connected by windows, each with a characteristic radius drawn from a distribution \citep{Mason1988}. Parlar and Yortsos \citep{Parlar1988} extended this model to more completely describe the case of scanning isotherms. Seaton \citep{Seaton1991} applied the model to the calculation of the average coordination number of pores in various materials, while Tanev and Vlaev \citep{Tanev1993} considered the relationship between assumptions about the pore shape and the corresponding hysteresis loop.

The models presented in the above articles focus primarily on the percolation threshold. They assume that no emptying of a significant number of pores occurs until the vapor region is able to percolate through this network. This assumption seems to work well for materials where the pore size distribution is not too wide, such as porous glass \citep{Mason1988}. An interesting and not fully addressed scenario is when the material, with a broad PSD, has a fraction of large pores empty at the start of desorption, immediately below the saturation pressure. This could be because these pores are so large that they do not fill until relative humidity approaches very close to 1 in the experiments, or because they are able to empty without hindrance due to a connection to a continuous network of large pores, in which case evaporation can be much more efficient due to the "small world" effect~\citep{moore2000epidemics, watts1998collective, latora2001efficient}.

Since the network provides a hindrance only to drying, the sorbate content during wetting is the same as that in the independent pore model given in Eqn.~\ref{adsorbingmass}. The desorption branch can be calculated by mean-field percolation models~\citep{Mason1988, liu1993analysis, Liu1994, neimark1991percolation}, summarized as following.
Assuming cylindrical pores, at a particular relative pressure $h$, the fraction $q$ of pores that are below their emptying transition pressure, referred to as a ``large pore", is given by
\begin{equation}
q = \frac{\int_{r_{eq}(h)}^\infty\frac{v(r)}{r^2}~dr}{\int_0^\infty\frac{v(r)}{r^2}~dr}.
\end{equation}
The total sorbed volume during drying will then be
\begin{equation}
m_d(h) = \int_0^{r_{eq}(h)}v(r)~dr+(1-Q)\int_{r_{eq}(h)}^\infty v(r)~dr+Q\int_{r_{eq}(h)}^\infty\frac{r^2-[r-at_m(h)]^2}{r^2}v(r)~dr, \label{desorbnospinodal}
\end{equation}
where Q is the probability of such a large pore being actually empty.
Various lattices can be chosen such as simple cubic~\citep{liu1993analysis} or a Bethe lattice with coordination number  $z$ ~\citep{Mason1983,Seaton1991} and Q can be calculated based on the lattice choice. 

\begin{figure*}
\centering
\includegraphics[width=\textwidth]{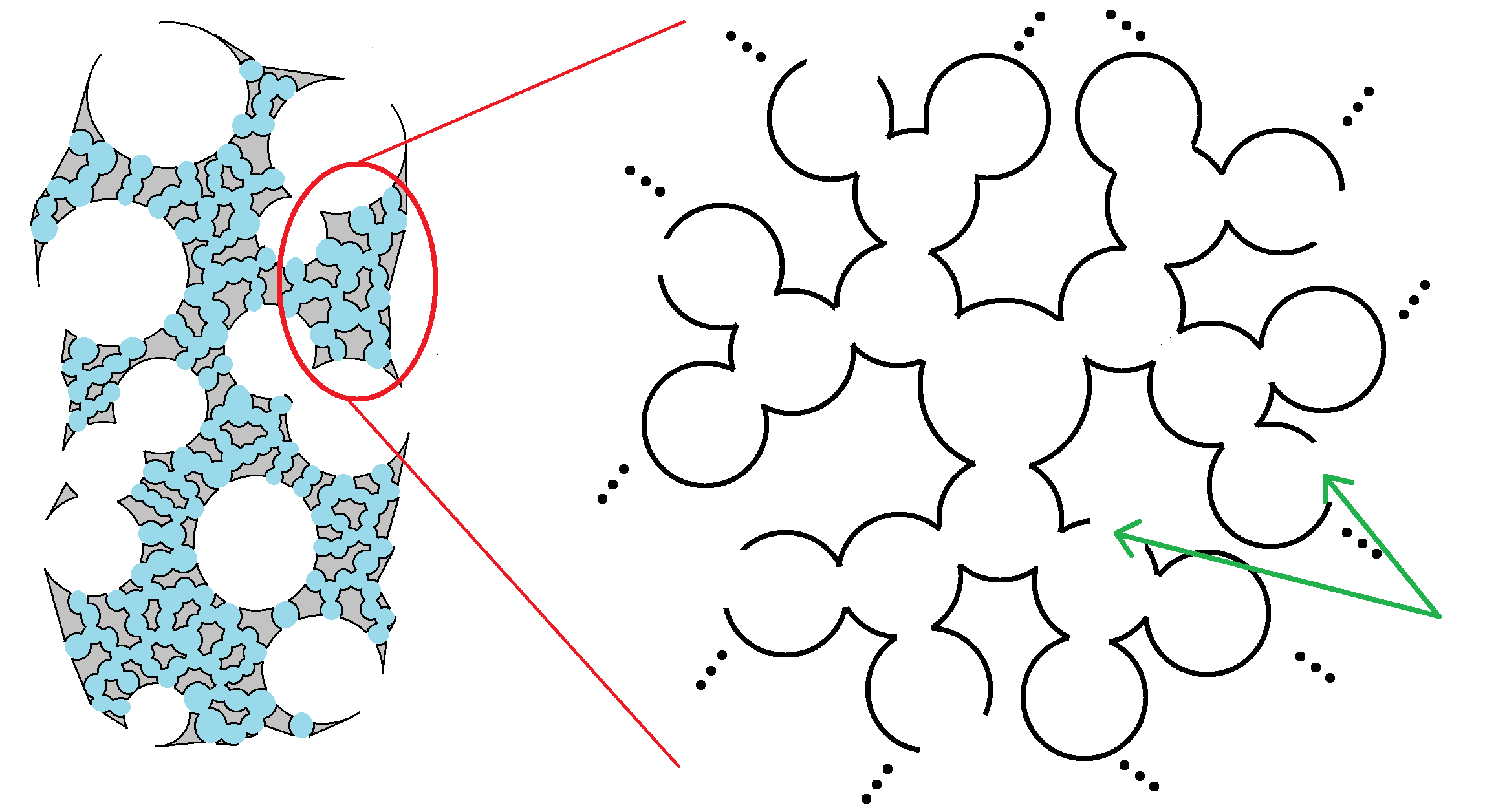}
\caption{A schematic diagram of a hierarchical network of small (blue) and larger always exposed (white) pores formed by a solid skeleton (grey), and its representation by a modified Bethe lattice, visualized with $z = 3$. The holes in some of the displayed pores, noted by green arrows, represent an interface between the network of small pores and larger pores that are always exposed and acting as a finite boundary to the originally infite Bethe lattice. \label{fig:network}}
\end{figure*}

However, no single infinite-lattice model without finite size effects can adequately describe both simple pore structures such as those of Vycor glass or silica gel, and more hierarchical pore networks such as in cement paste or dental enamel. We extend and generalize the percolation models by introducing a parameter $f$, defined to be the fraction of all pores exposed to a liquid-vapor interface as soon as drying commences, so that the pores empty as soon as their equilibrium transition pressure is reached. The exposure may, for instance, be due to a connection to a very large percolating pore that remains empty, as illustrated in Fig. \ref{fig:network}, and these always exposed pores bring in the boundary effect of an effectively finite lattice. There are thus two mechanisms by which a particular pore may have access to the vapor phase. The first is direct exposure, with probability $f$. The second is that it could be connected to the vapor by a neighbor on the lattice. We denote $X$ to be the probability that an individual neighbor provides such a connection. We then can write
\begin{equation}
1-Q = (1-f)(1-X)^z. \label{percolation}
\end{equation}
Due to the self-similarity of the Bethe lattice, we can further write
\begin{equation}
X = q[f+(1-f)(1-(1-X)^{z-1})]. \label{selfc}
\end{equation}
We solve equation \ref{selfc} self-consistently to find $X$, and then use equation \ref{percolation} to calculate $Q$.

\section{Cavitation}
One further effect controls desorption at low relative pressure. This effect has been called cavitation \citep{Thommes2006, thommes2015physisorption} or the tensile strength effect \citep{Groen2003}. The cause of desorption is that the condensed fluid becomes unstable at sufficiently low pressure, without direct exposure to a vapor phase.

We assume that the primary result of this desorption mechanism is the homogeneous nucleation of the vapor phase in blocked large pores. The pressure at which confined cavitation~\citep{rasmussen2010cavitation} or spinodal decomposition~\citep{hildebrand1927quantitative, bazant2013theory} occurs in a pore of width $r_n$ can be derived as
\begin{equation}
h_n(r_n) = h_n(\infty) e^{-\frac{2\gamma v}{[r_n-at(h_n)]kT}},
\end{equation}
where $h_n(\infty)$ is a parameter in the model, with a value that depends on the sorbate and temperature but not on the sorbent. This gives a relationship between relative pressure and the width above which all pores must be empty. If we define
\begin{equation}
q_n = \frac{\int_{r_n}^\infty\frac{v(r)}{r^2}~dr}{\int_0^\infty\frac{v(r)}{r^2}~dr}, \label{equationqn}
\end{equation}
we can then write
\begin{equation}
X = q_n+(q-q_n)[f+(1-f)(1-(1-X)^{z-1})]
\end{equation}
and
\begin{equation}
\begin{split}
m_d(h) = & \int_0^{r_{eq}(h)}v(r)~dr+(1-Q)\int_{r_{eq}(h)}^{r_n}v(r)~dl \\
& + Q\int_{r_{eq}(h)}^{r_n}\frac{r^2-[r-at_m(h)]^2}{r^2}v(r)~dr 
 +\int_{r_n}^\infty\frac{r^2-[r-at_m(h)]^2}{r^2}v(r)~dr.
\end{split}
\label{eqn:extended-desorbingmass}
\end{equation}

\section{Results of the extended finite-size percolation model}
\subsection{Adsorption/desorption isotherms and values for f}
\begin{figure}
\centering
\begin{subfigure}{0.78\textwidth}
\centering
\includegraphics[width=0.98\textwidth]{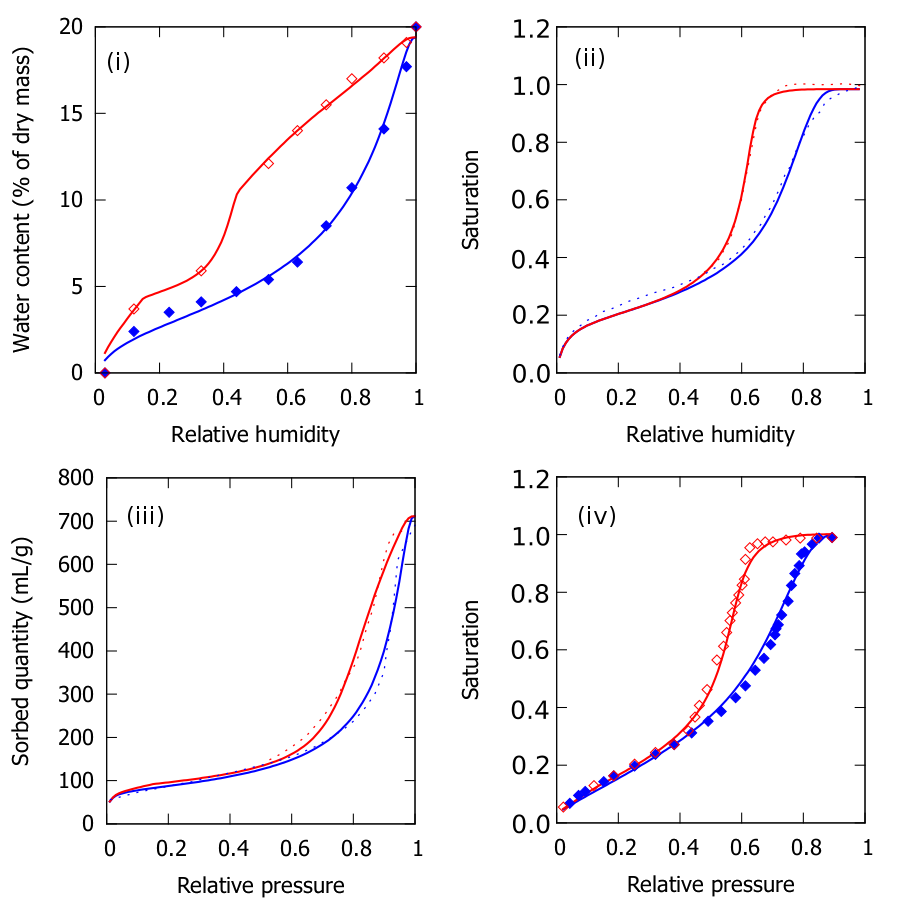}
\subcaption{\label{fig:result-examples}}
\end{subfigure}
%\\
\caption{Extended model applied on various materials. Calculated sorption isotherms (lines), along with experimental isotherms (points or dashed lines) for wetting (blue) and drying (red) of (i) water in hardened cement paste \citep{Baroghel-Bouny2007}, (ii) water in Vycor \citep{Li1991}, (iii) nitrogen in carbon nanotubes \citep{Frackowiak2001} and (iv) xenon in silica glass \citep{Machin1994}.}
\end{figure}

\begin{figure}
\centering
\begin{subfigure}{0.48\textwidth}
\centering
\includegraphics[width=0.98\textwidth]{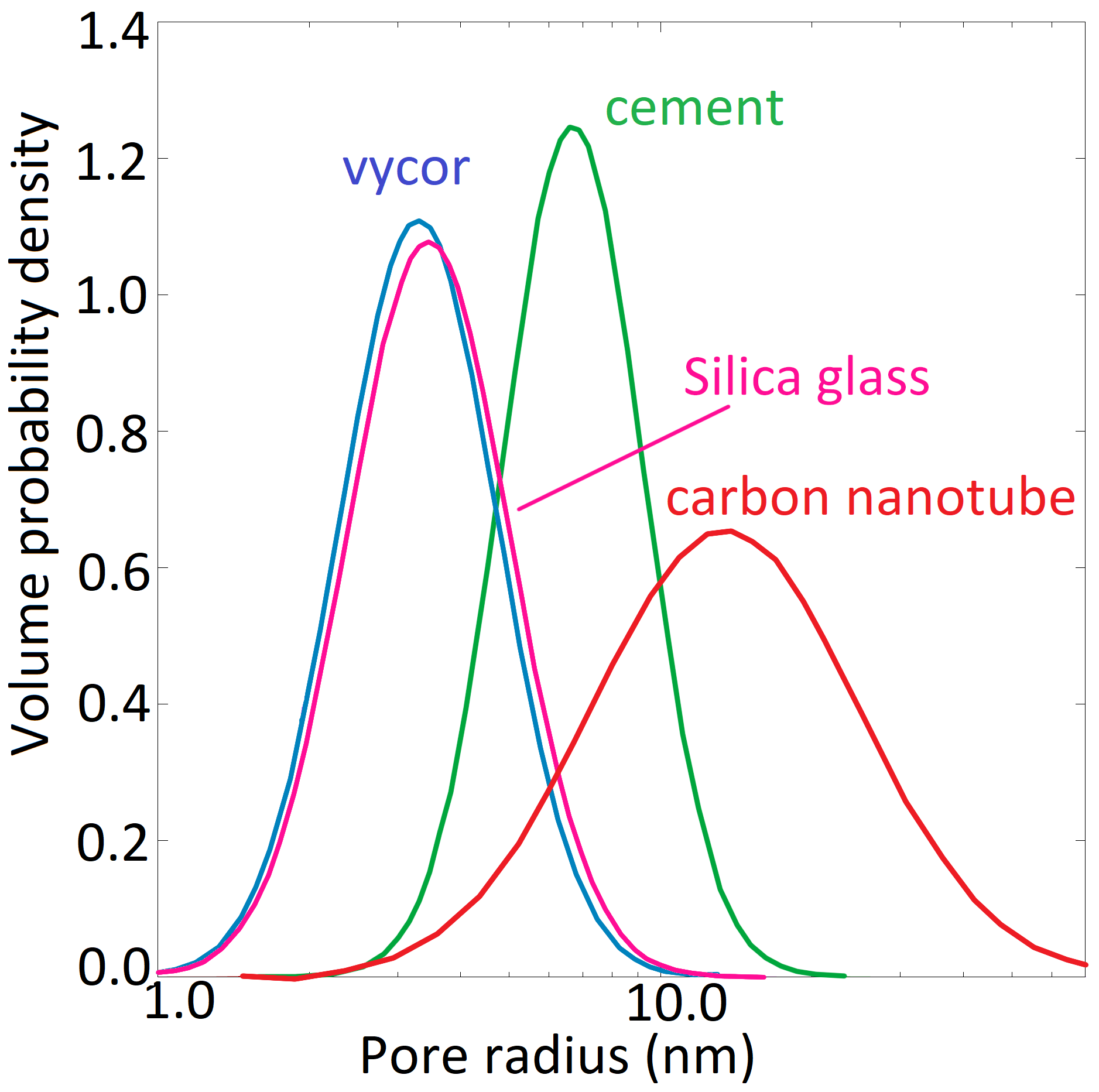}
\subcaption{\label{fig:result-psd}}
\end{subfigure}
\begin{subfigure}{0.48\textwidth}
\includegraphics[width=0.98\textwidth]{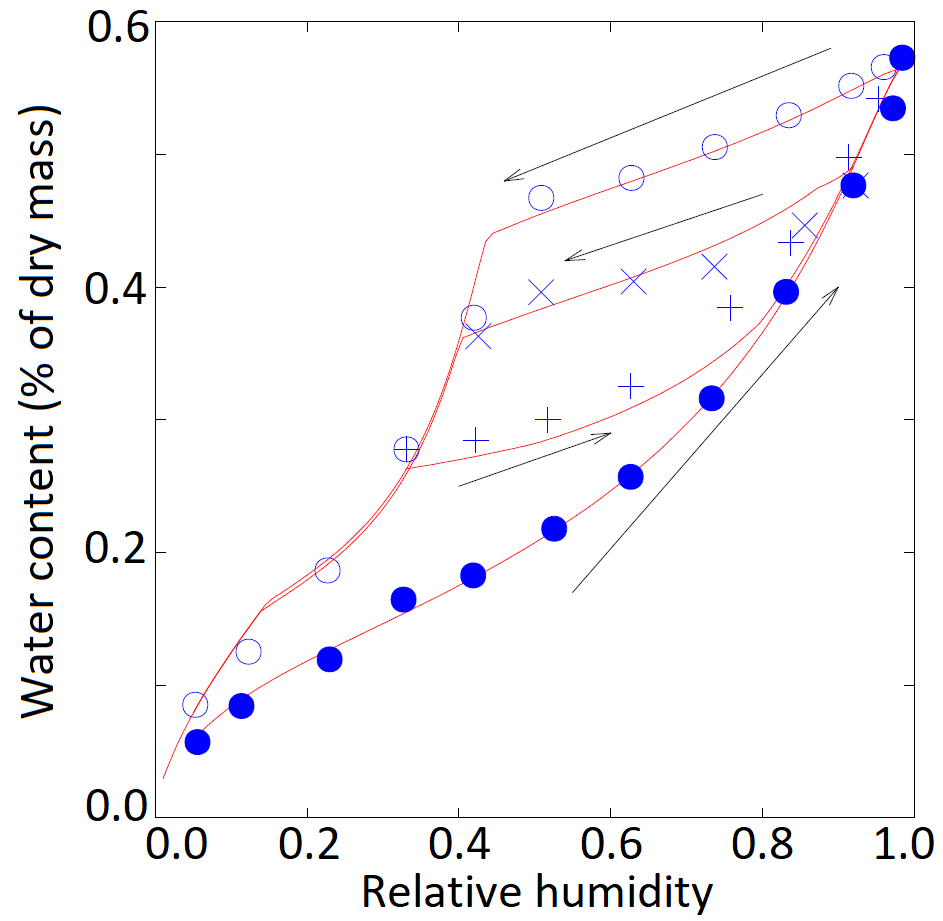}
\subcaption{\label{fig:result-scanningisotherm}}
\end{subfigure}
\caption{Extended model applied on various materials. (a) Volume density function for hardened cement paste \citep{Baroghel-Bouny2007}, Vycor \citep{Li1991}, carbon nanotubes \citep{Frackowiak2001} and silica glass \citep{Machin1994}. (b) Primary and scanning sorption isotherms calculated using the model of this work (solid lines), along with experimental primary (circles) and scanning (crosses) sorption isotherms for water in dental enamel at 298 K \citep{Zahradnik1975}.}
\end{figure}

Figure \ref{fig:result-examples} compares our extended model results with experimental observations for various systems \citep{Baroghel-Bouny2007,Frackowiak2001,Li1991,Machin1994}. The model parameters were determined in each case by a least squares fit. \change{To avoid overfitting, we fix the value of $z=4$ throughout this study for all materials concerned, focusing on the new order parameter $f$ proposed here.} Figure \ref{fig:result-psd} shows the calculated pore size distributions for these materials. The median radius of 13 nm found for the case of carbon nanotubes is consistent with the transmission electron microscope image of these nanotubes (figure 12 of \cite{Frackowiak2001}).

Values of $f$ calculated for various materials are summarized in table \ref{ftable}. Uncertainty ranges are calculated by comparing isotherms of samples produced with different specifications (hardened cement paste), or taken at different temperatures (dental enamel) or using different sorbates (porous silica glass and Vycor). This is statistical uncertainty only and does not account for model uncertainty. The calculated $f$ values indicate that the finite size effect of percolation lattice model is \change{more significant for some hierarchical porous materials such as cement paste and enamel than other materials such as Vycor and silica glass. The percolating large pores in these systems with higher $f$ values are expected to have prominent} effects on the sorption isotherm, consistent with other studies~\citep{pinson2015hysteresis, ioannidou2016mesoscale}. Further direct measurements of the $f$ values in these systems will be very interesting to compare with the presented model.

\begin{table}
\centering
\caption{Fraction of always exposed pores, $f$, calculated by applying the model presented here to a variety of published experimental data. \label{ftable}}
\begin{tabular}{|cccc|}
\hline
\textbf{Material} & \textbf{Sorbate/s} & \textbf{Reference} & \textbf{f} \\
\hline
CNT & \ce{N_2} & \citep{Frackowiak2001} & 0.40 \\
\hline
Cement & \ce{H_2O} & \citep{Feldman1968a} \citep{Baroghel-Bouny2007}
 & $0.33\pm 0.06$ \\
\hline
Enamel & \ce{H_2O} & \citep{Zahradnik1975} & $0.54\pm 0.05$ \\
\hline
Silica glass & 
\begin{tabular}{@{}c@{}}
Xe, \ce{C_4H_{10}} \\
\ce{CCl_2F_2}, \ce{O_2}
\end{tabular}
& \citep{Machin1994} & $0.13\pm 0.09$ \\
\hline
 & \ce{N_2} & \citep{Brewer1962} & \\
Vycor & \ce{H_2O} & \citep{Li1991} & $0.06\pm 0.01 $ \\
 & \ce{C_6H_{14}} & \citep{Page1995}
 & \\
\hline
\end{tabular}
\end{table}

\subsection{Scanning isotherms}
The principles of the model can be applied directly to the case of scanning isotherms, where the partial pressure is cycled over some subsection of the possible range.

If the decrease in partial pressure is halted at $h_-$ and the partial pressure then increased to $h$, the total sorbed volume will be
\begin{align*}
m_{\uparrow scan}(h) & = \int_0^{r_K(h_-)+at(h_-)}v(r)~dr+[1-Q(h_-)]\int_{r_K(h_-)+at(h_-)}^{r_n(h_-)}v(r)~dr \\
& +Q(h_-)\int_{r_K(h_-)+at(h_-)}^{r_n(h_-)}\frac{r^2-[r-at_m(h)]^2}{r^2}v(r)~dr \\
& +\int_{r_n(h_-)}^\infty\frac{r^2-[r-at_m(h)]^2}{r^2}v(r)~dr,
\end{align*}
where any non-zero integral bound must be replaced by $r_{fill}(h)$ when the latter surpasses the former.

If the increase in partial pressure is halted at $h_+$ and the partial pressure then decreased to $h$, the total sorbed mass can be calculated using equations \ref{equationqn} and \ref{desorbingmass}, but using the smaller of $r_n(h)$ and $r_K(h_+)/2+at(h_+)$ in place of $r_n(h)$, i.e. the radius of the largest pores that could be filled.

Figure \ref{fig:result-scanningisotherm} compares the model with experimental scanning isotherms for dental enamel \citep{Zahradnik1975}.

\begin{figure}
\centering
\includegraphics[width=0.78\textwidth]{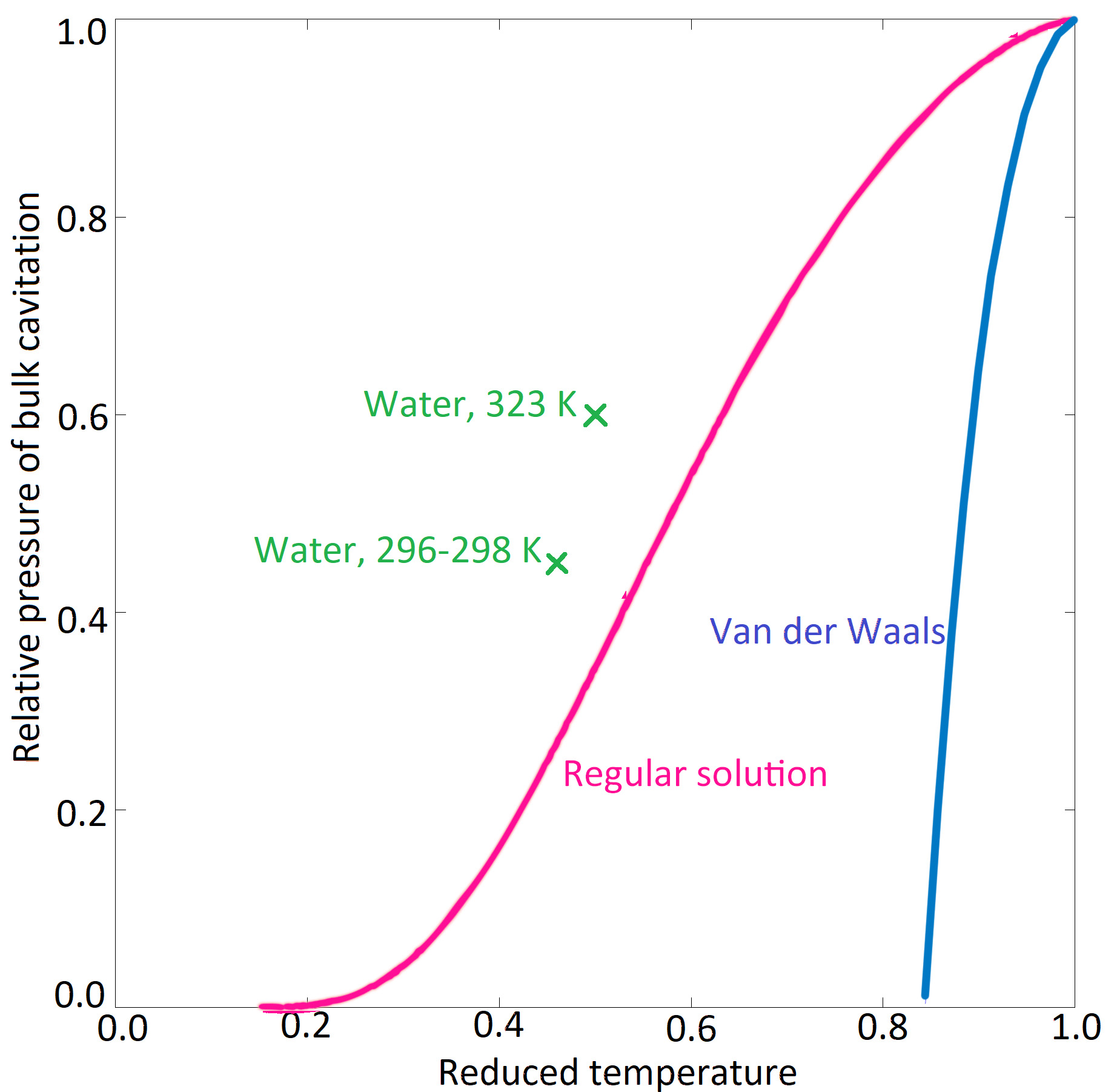}
\caption{Partial pressure at which the vapor phase is nucleated in the bulk, as a function of reduced temperature. Points denote values obtained by applying the model of this work to experimental data, while the solid lines are the results of the regular solution and van der Waals models. \label{fig:cavitation}}
\end{figure}

\change{
\subsection{Prediction of liquid cavitation pressure}
Liquid cavitation and fracture under tensile stress has long been a topic of interest~\citep{fisher1948fracture, trevena1987cavitation, oxtoby1992homogeneous, zheng1991liquids, levitas2011phase}. One tantalising prospect is the use of experiments like those cited in this work to measure the bulk cavitation relative pressure $h_n(\infty)$ for various fluids as a function of temperature. In only two distinct cases, water around 25~{\C} and at 50~{\C}, was the characteristic knee observed. $h_n(\infty)$ was larger at higher temperature as expected. A systematic study could allow this quantity to be measured as a function of temperature, providing insight into the nature of forces between molecules in a liquid. For example, the regular solution model~\citep{hildebrand1927quantitative} predicts
\begin{equation}
h_\infty = \frac{e^{-2\sqrt{1-\tau}/\tau}(1+\sqrt{1-\tau})}{1-\sqrt{1-\tau}},
\end{equation}
where $\tau = T/T_c$ is temperature divided by the critical temperature of the sorbate (reduced temperature). Likewise, a universal curve can be obtained for a Van der Waals gas, though there is no simple analytic form. The independence of the curves on material specifics suggests that there is a general trend across sorbates. Figure \ref{fig:cavitation} compares available points found using the model, in which $h_n(\infty)$ depends on sorbate and temperature but not sorbent, with the theoretical curves of regular solution model and Van der Waals gas.
}

\section{Discussion}
\change{
Based on our analysis of physical mechanisms for sorption hysteresis, we present a theoretical modeling approach to investigate pore connectivity from sorption hysteresis experiments. 
The model presented provides an overall description of hysteresis sorption in mesoporous media, especially in materials in which macropores are also present. The existence and connectivity of these macropores are characterized by an parameter $f$ describing the prevalence of large pores, in addition to the average pore coordination number  $z$. We calculate pore size distribution and connectivity for various materials, predict scanning isotherms and compare them to experiment data. Bulk liquid cavitation pressure can be inferred as a parameter in our model controlling the existence and position of the desorption knee. We find the bulk cavitation relative pressure for water increases with increasing temperature, using the best fit values from our model calculation on water sorption data. Theoretical curves given by the regular solution model and Van der Waals liquid predict a similar trend, with the regular solution model quantitatively closer to our results. More direct experiment data on bulk liquid rupture will help validate the models.
}

\change{
Nevertheless, there remain some complications that increase the uncertainty in the level to which the model can quantitatively describe pore size and network structure. 
First, the model as formulated assumes cylindrical pores, and a particular amount of single-pore hysteresis due to differences in the shape of the liquid-vapor interface. Other assumptions, such as slit pores with larger hysteresis or equilibrium filling and hence no hysteresis, could have been made instead and built into the model in a straightforward way. Agreement with TEM observations for the case of carbon nanotubes gives some confidence that the assumption of cylindrical pores is reasonable. However, the assumption still influences the obtained results. Specifically, if cylindrical pores are assumed when the system is really made up of slit pores, the obtained pore sizes will be larger and obtained value of $f$ smaller than in reality (the discrepancies are opposite if in fact the pores are able to reach local equilibrium on filling).
Second, the model uses a very simple model of surface adsorption. The effect of this can be seen in some discrepancy between model and experimental results at low pressure. Where the data are available, this could be rectified by using experimental t curves, at the loss of some generality. 
Third, the assumption of a fixed value of $z$ precludes distinguishing between systems based on their connectivity. For example, we might expect carbon nanotubes to have $z \approx 2$, while disordered porous systems such as glasses and cement paste have higher  $z$. Sorption data alone are not sufficient to allow such an identification to be made. Also,
it seems to be $f$ rather than  $z$  that is most influential in determining the sorption behavior.
}

The modeling approach here has been used in previous studies on cement sorption, drying shrinkage and creep~\citep{pinson2015hysteresis, masoero10modelling, jennings2013water}, where the multiscale porous structure of cement paste exemplifies a high $f$ value, and a distinct bump on the desorption branch at around 40~\% relative humidity exists due to cavitation. The good comparison with cement sorption and shrinkage data demonstrates the power of this physics based theoretical model.

The model has the potential to contribute to transport modeling not only through its assessment of connectivity, but also by identifying and explaining the hysteresis in the sorbed mass as a function of local sorbate chemical potential. Existing models of transport in porous materials do not account for this hysteresis, simplified by defining an effective diffusivity $D(S)$ \citep{Martys1997,Baroghel-Bouny2007a} by
\begin{equation}
\frac{\partial S}{\partial t} = \nabla.[D(S)\nabla S],
\end{equation}
where $S$ is the local saturation of sorbate. Although this can give a qualitative idea of the conditions under which transport is faster or slower, more accurate transport rate can be determined with knowledge of local chemical potential gradients, which can be calculated though a model such as the one presented here. Deeper understanding of the transport processes will assist in the design of these materials with desired transport properties.  The transport of water and of ions dissolved in the water is also highly relevant to degradation and contamination in concrete \citep{Faucon1998,Samson2000,Glasser2008} and many other materials \citep{Simunek1994,Mayer2002}.

\section{Conclusions}
The hysteresis in adsorption/desorption isotherms caused by the collective effect within a pore network has been mainly modeled by idealized infinite lattice percolation models\citep{Mason1982, Mason1983, Mason1988, liu1993analysis, Liu1994, neimark1991percolation}. 
We present an extension to these percolation models by introducing a parameter $f$ characterizing the fraction of always exposed large pores, effectively incorporating boundary effects of percolation on a finite size lattice. Our calculated $f$ values indicate the importance of finite size percolation effects in different materials. 
\change{
The model is simplified with certain assumptions as discussed above, nevertheless it addresses the currently missing link between 3D simulations of colloidal systems and macroscopic continuum models of vapor sorption, and provides a pathway to overcome the current empirical treatment of sorption isotherms in continuum poromechanical models \citep{masoero2018c,gawin1999numerical,davie2010fully,di2009hygro}.
}

The significant bump in the desorption isotherm, which typically appears in multiscale materials such as cement paste,  around 40\% relative pressure in the case of water at 25 $^\circ$C, can be well explained by an effect of cavitation (equivalently, homogeneous nucleation). The pressure at which this bump is located in experimental data varies with temperature in a trend consistent with theoretical expectations, though available data are limited.

The development here of an extended model relating macroscopic sorption behavior to microstructure can assist the design of materials for applications such as moisture buffering \citep{Osanyintola2006}, and the understanding of the dependence of thermal conductivity on water content \citep{Jerman2012}.
Since connectivity plays a vital role in determining the transport properties of a material \citep{Hall1987,Hall1995}, it would be very interesting for future research to examine the relationship between $f$ and  $z$ and the permeability to single-phase flow. \change{  More generally, for multiphase flow in porous media, this could lead to improved mathematical models of the relative permeability and capillary pressure~\cite{gu2018microscopic}, which also exhibit strong and poorly understood hysteresis.}

\vspace{0.1in}

This work was supported by the Concrete Sustainability Hub at MIT with sponsorship provided by the Portland Cement Association (PCA) and the Ready Mixed Concrete (RMC) Research \& Education Foundation, and by the National Science Foundation via an EAGER Collaborative Grant \#1153509. The authors acknowledge useful discussions with the members of the Dome collaboration and G. Sant. 





%\begin{thebibliography}{00}


%\end{thebibliography}

\bibliographystyle{elsarticle-num} 
\bibliography{library}

\end{document}